\documentclass[aps,pra,twocolumn,groupedaddress,showpacs]{revtex4-1}
\usepackage{amsmath,bm}
\usepackage[dvipdfmx]{graphicx,color}
\newcommand{\oriham}{\mathcal{H}_0(t)}
\newcommand{\beq}{\begin{equation}}
\newcommand{\eeq}{\end{equation}}
\newcommand{\beqs}{\begin{eqnarray}}
\newcommand{\eeqs}{\end{eqnarray}}
\newcommand{\tder}{\frac{\partial}{\partial t}}
\begin{document}
\title{Shortcuts to adiabaticity in the infinite-range Ising model \\ by mean-field counter-diabatic driving}
\author{Takuya Hatomura}
\email[]{hatomura@spin.phys.s.u-tokyo.ac.jp}
\affiliation{Department of Physics, Graduate School of Science, The University of Tokyo, 7-3-1 Hongo, Bunkyo, Tokyo, Japan}
\date{\today}
\begin{abstract}
The strategy of shortcuts to adiabaticity enables us to realize adiabatic dynamics in finite time. 
In the counter-diabatic driving approach, an auxiliary Hamiltonian which is called the counter-diabatic Hamiltonian is appended to an original Hamiltonian to cancel out diabatic transitions. 
The counter-diabatic Hamiltonian is constructed by using the eigenstates of the original Hamiltonian. 
Therefore, it is in general difficult to construct the counter-diabatic Hamiltonian for quantum many-body systems. 
Even if the counter-diabatic Hamiltonian for quantum many-body systems is obtained, it is generally non-local and even diverges at critical points. 
We construct an approximated counter-diabatic Hamiltonian for the infinite-range Ising model by making use of the mean-field approximation. 
An advantage of this method is that the mean-field counter-diabatic Hamiltonian is constructed by only local operators. 
We numerically demonstrate the effectiveness of this method through quantum annealing processes going the vicinity of the critical point. 
It is also confirmed that the mean-field counter-diabatic Hamiltonian is still well-defined in the limit to the critical point. 
The present method can take higher order contributions into account and is consistent with the variational approach for local counter-diabatic driving. 
\end{abstract}
\pacs{64.60.Ht,05.30.Rt,64.70.Tg,75.40.Mg}
\maketitle
%
%
\section{\label{Sec.Intro}Introduction}
Shortcuts to adiabaticity are collectively called methods which enable us to realize adiabatic dynamics in finite time~\cite{doi:10.1021/jp030708a,doi:10.1021/jp040647w,doi:10.1063/1.2992152,1751-8121-42-36-365303,PhysRevLett.104.063002,Torrontegui2013117}. 
One of the famous methods is the counter-diabatic driving approach~\cite{doi:10.1021/jp030708a,doi:10.1021/jp040647w,doi:10.1063/1.2992152,1751-8121-42-36-365303} (or in other words, the transitionless quantum driving approach), in which an additional auxiliary Hamiltonian is introduced to let a system be adiabatic in an original Hamiltonian frame. 
In contrast, in the invariant-based inverse engineering approach~\cite{PhysRevLett.104.063002}, variables of a Hamiltonian are parametrized according to schedules of a dynamical invariant~\cite{doi:10.1063/1.1664991}. 
At first, these methods have been applied to simple systems which have small degrees of freedom~\cite{Torrontegui2013117}. 

Application of these methods to quantum many-body systems is a curious problem and recently has been investigated as a central issue~\cite{PhysRevLett.109.115703,PhysRevE.87.062117,PhysRevA.90.060301,1742-5468-2014-12-P12019,PhysRevLett.114.177206,1607.05687,PhysRevA.95.012309}. 
In quantum many-body systems, quantum phase transitions drastically change properties of systems~\cite{Sachdev,Nishimori}. 
In particular, when we consider quenched dynamics, critical points of systems play an important role. 
That is, correlation length and relaxation time diverge at critical points, which leads to effectively frozen dynamics~\cite{0305-4470-9-8-029,Kibble1980183,Zurek1985,Zurek1993,doi:10.1080/00018732.2010.514702,RevModPhys.83.863,doi:10.1142/S0217751X1430018X}. 
It means that adiabatic dynamics is broken down around critical points. 
Associated with the divergence of correlation length, the counter-diabatic Hamiltonian for quantum many-body systems becomes non-local and even diverges at critical points. 
In addition, the counter-diabatic driving approach requires the eigenstates of a Hamiltonian. 
Therefore, application to general quantum many-body systems is quite restricted. 

In infinite-dimensional system cases, the infinite-range Ising model, which is also known as the Lipkin-Meshkov-Glick model, was studied~\cite{PhysRevE.87.062117,PhysRevLett.114.177206,PhysRevA.95.012309}. 
One of the ways to treat this model is taking the Holstein-Primakoff transformation in the thermodynamic limit~\cite{PhysRevE.87.062117,PhysRevLett.114.177206}. 
Then, this model is mapped to a harmonic oscillator and an approximated counter-diabatic Hamiltonian can be obtained. 
In combination with the optimization approach, quasi-adiabatic dynamics was realized across the critical point~\cite{PhysRevLett.114.177206}. 
This model was also investigated in the invariant-based inverse engineering approach~\cite{PhysRevA.95.012309}. 
Although dynamical invariants of this model generally contain infinitely non-local operators, a local dynamical invariant was constructed by making use of the mean-field approximation. 
The similar idea should be also explored in the counter-diabatic driving approach. 

In this paper, we construct a local counter-diabatic Hamiltonian for the infinite-range Ising model by making use of the mean-field approximation. 
Then, we can express the counter-diabatic Hamiltonian by only local operators although there is an auxiliary self-consistent equation. 
Following the previous work~\cite{PhysRevA.95.012309}, we numerically demonstrate the usefulness of this method through quantum annealing processes. 
We also show the non-divergence of the mean-field counter-diabatic Hamiltonian in the limit to the critical point for certain situations. 
The relation to the variational approach for local counter-diabatic driving~\cite{1607.05687} is also of interest. 
We confirm that the present mean-field counter-diabatic driving approach includes higher order contributions, which cannot be taken into account if we naively adopt the variational approach. 

This article is constructed as follows. 
Section~\ref{Sec.STA} is devoted to the brief review of shortcuts to adiabaticity by the counter-diabatic driving approach and the simplest example is shown as a preliminary in Sec.~\ref{Sec.twolevel}. 
In Sec.~\ref{Sec.Example}, the couter-diabatic Hamiltonian for the infinite-range Ising model is constructed by making use of the mean-field approximation. 
The infinite-range Ising model can be easily calculated by numerical simulations because it is equivalent to the uniaxial single-spin system as seen in Sec.~\ref{Sec.numsim}. 
The usefulness of the mean-field counter-diabatic driving is demonstrated through quantum annealing processes in Sec.~\ref{Sec.annealing}. 
The non-divergence of the mean-field counter-diabatic Hamiltonian in the limit to the critical point is shown for a certain class of schedules of the transverse field in Sec.~\ref{Sec.explicit}. 
The relation to the variational approach for local counter-diabatic driving is discussed in Sec.~\ref{Sec.local}. 
We summarize in Sec.~\ref{Sec.Summary}. 

%
%
\section{\label{Sec.Method}Method}
%
%
\subsection{\label{Sec.STA}Counter-diabatic driving}
We consider a time-dependent Hamiltonian $\oriham$ with the eigenenergies and the eigenstates
\beq
\oriham|\psi_n(t)\rangle=E_n(t)|\psi_n(t)\rangle.
\label{Eq.Eigeneq}
\eeq
The time-evolution operator of adiabatic dynamics is defined by
\beq
U(t)=\sum_ne^{i\alpha_n(t)}|\psi_n(t)\rangle\langle\psi_n(0)|,
\label{Eq.Evolve}
\eeq
where $\alpha_n(t)$ is the dynamical phase
\beqs
\alpha_n(t)&=&-\frac{1}{\hbar}\int_0^tdt^\prime E_n(t^\prime) \notag \\
&&+i\int_0^tdt^\prime\langle\psi_n(t^\prime)|\partial_{t^\prime}\psi_n(t^\prime)\rangle.
\label{Eq.Dynaphase}
\eeqs
We construct the assisted Hamiltonian $\mathcal{H}(t)$ so that the adiabatic dynamcis $|\Psi(t)\rangle=U(t)|\Psi(0)\rangle$ becomes the solution of the Schr\"odinger equation 
\beqs
i\hbar\tder|\Psi(t)\rangle&=&\mathcal{H}(t)|\Psi(t)\rangle, \label{Eq.Schro} \\
|\Psi(t)\rangle&=&\sum_nc_ne^{i\alpha_n(t)}|\psi_n(t)\rangle, \label{Eq.Adstate}
\eeqs
where $c_n$ is the coefficient of the initial state, $|\Psi(0)\rangle=\sum_nc_n|\psi_n(0)\rangle$. 
By inversely solving Eq.~(\ref{Eq.Schro}), we find that the assisted Hamiltonian is constructed as
\beqs
\mathcal{H}(t)&=&\oriham+\mathcal{H}_\mathrm{cd}(t), \label{Eq.Totham} \\
\mathcal{H}_\mathrm{cd}(t)&=&i\hbar\sum_{n\neq m}\langle\psi_n(t)|\partial_t\psi_m(t)\rangle|\psi_n(t)\rangle\langle\psi_m(t)|. \label{Eq.CDdriving}
\eeqs
where $\mathcal{H}_\mathrm{cd}(t)$ is called the counter-diabatic Hamiltonian. 

In order to construct the counter-diabatic Hamiltonian, the eigenstates of the Hamiltonian $\oriham$ are required as seen in Eq.~(\ref{Eq.CDdriving}). 
However, in general, it is difficult to explicitly obtain the eigenstates of quantum many-body systems. 
Even if the eigenstates are obtained, the counter-diabatic Hamiltonian becomes highly non-local and even diverges at critical points. 

%
%
\subsection{\label{Sec.twolevel}Preliminary: Two-level system}
When we adopt the mean-field approximation, problems of the infinite-range Ising model is reduced to problems of a two-level system. 
Therefore, as a preliminary, we consider a two-level system
\beq
\oriham=-\Gamma(t)\sigma^x-h(t)\sigma^z,
\label{Eq.Twolevelham}
\eeq
where $\sigma^\alpha, \alpha=x,y,z$ are the Pauli matrices, $\Gamma(t)$ is a transverse field, and $h(t)$ is a longitudinal field. 
The eigenenergies and the eigenstates are given by
\beqs
E_\pm(t)&=&\pm\sqrt{h(t)^2+\Gamma(t)^2}, \label{Eq.Twolevelene} \\
|\psi_-(t)\rangle&=&
\begin{pmatrix}
\cos\theta \\
\sin\theta
\end{pmatrix}
,\quad|\psi_+(t)\rangle=
\begin{pmatrix}
-\sin\theta \\
\cos\theta
\end{pmatrix}
, \label{Eq.Twolevelstate}
\eeqs
with
\beqs
\sin2\theta&=&\frac{\Gamma(t)}{\sqrt{h(t)^2+\Gamma(t)^2}}, \\
\cos2\theta&=&\frac{h(t)}{\sqrt{h(t)^2+\Gamma(t)^2}}. 
\eeqs
The counter-diabatic Hamiltonian is given by
\beqs
\mathcal{H}_\mathrm{cd}(t)&=&\dot{\theta}(t)\sigma^y, \\
\dot{\theta}(t)&=&\frac{1}{2}\frac{h(t)\dot{\Gamma}(t)-\dot{h}(t)\Gamma(t)}{h^2(t)+\Gamma^2(t)}. 
\eeqs
Here and hereafter, $\hbar=1$. 
%
%
\subsection{\label{Sec.Example}Mean-field counter-diabatic driving \\ for the infinite-range Ising model}
The Hamiltonian of the infinite-range Ising model is given by
\beq
\mathcal{H}_0(t)=-\frac{J}{2N}\sum_{i,j}\sigma_i^z\sigma_j^z-\Gamma(t)\sum_i\sigma_i^x-h\sum_i\sigma_i^z,
\label{Eq.orihamlong}
\eeq
where $J$ is the coupling constant and $N$ is the number of Ising spins. 
Throughout this paper, we consider the time-dependent transverse field $\Gamma(t)$ and the fixed longitudinal field $h$. 
There exists the critical point at the transverse field $\Gamma(t)=J$ and the longitudinal field $h=0$. 
The system is in the ferromagnetic phase for $\Gamma(t)<J$, and the paramagnetic phase for $\Gamma(t)>J$. 

Owing to the long-range nature of the interactions, the mean-field approximation is valid for static properties of large $N$ systems. 
By using the mean-field approximation, we obtain the following mean-field Hamiltonian
\beqs
\mathcal{H}_0^\mathrm{MF}(t)&=&\frac{JN}{2}(m^z(t))^2-(Jm^z(t)+h)\sum_i\sigma_i^z \notag \\
&&-\Gamma(t)\sum_i\sigma_i^x,
\label{Eq.LongMFham}
\eeqs
where $m^z(t)$ is given by the expectation value of $\sigma_i^z$ and determined later. 
In the mean-field Hamiltonian (\ref{Eq.LongMFham}), $N$ spins are decoupled from each other. 
Therefore, the eigenstates are the products of the eigenstates of the two-level system (\ref{Eq.Twolevelstate}), and the eigenenergies are the summations of the eigenenergies of the two-level system (\ref{Eq.Twolevelene}). 
We remark that the longitudinal field is modulated as $h(t)\to Jm^z(t)+h$ due to the existence of the mean-field. 
The counter-diabatic Hamiltonian derived by the mean-field approximation is given by
\beqs
\mathcal{H}_\mathrm{cd}^\mathrm{MF}(t)&=&\dot{\theta}(t)\sum_i\sigma_i^y, \label{Eq.cdhamlong} \\
\dot{\theta}(t)&=&\frac{1}{2}\frac{(Jm^z(t)+h)\dot{\Gamma}(t)-J\dot{m}^z(t)\Gamma(t)}{(Jm^z(t)+h)^2+\Gamma^2(t)}.\quad\quad \label{Eq.cdfield}
\eeqs

In this paper, we only consider the ground state tracking. 
Therefore, we impose the following self-consistent equation
\beqs
m^z(t)&=&\langle\psi_-(t)|\sigma^z|\psi_-(t)\rangle \notag \\
&=&\frac{Jm^z(t)+h}{\sqrt{(Jm^z(t)+h)^2+\Gamma^2(t)}},
\label{Eq.SCE}
\eeqs
which can be rewritten as
\beqs
0&=&J^2(m^z(t))^4+2Jh(m^z(t))^3 \notag \\
&&-(J^2-h^2-\Gamma^2(t))(m^z(t))^2 \notag \\
&&-2Jhm^z(t)-h^2. \label{Eq.SCErewritten}
\eeqs
Here, this equation is a quartic equation with respect to the mean-field $m^z(t)$. 
Therefore, we can in principle solve this equation although the expression for the mean-field $m^z(t)$ is quite complicated. 
By differentiating Eq.~(\ref{Eq.SCErewritten}) with respect to time $t$, we obtain the time-derivative of the magnetization
\beqs
\dot{m}^z(t)&=&-\Gamma(t)\dot{\Gamma}(t)(m^z(t))^2 \notag \\
&&\times\{2J^2(m^z(t))^3+3Jh(m^z(t))^2 \notag \\
&&\quad-(J^2-h^2-\Gamma^2(t))m^z(t)-Jh\}^{-1}. \label{Eq.mzdot}
\eeqs 
In combination with Eqs.~(\ref{Eq.SCErewritten}) and (\ref{Eq.mzdot}), we can calculate the mean-field counter-diabatic field (\ref{Eq.cdfield}). 
%
%
\subsection{\label{Sec.numsim}Mapping to the single-spin system}
In order to confirm availability of the mean-field counter-diabatic Hamiltonian~(\ref{Eq.cdhamlong}), we perform numerical simulations. 
The total assisted Hamiltonian is given by the summation of (\ref{Eq.orihamlong}) and (\ref{Eq.cdhamlong})
\beqs
\mathcal{H}^\mathrm{MF}(t)&=&\oriham+\mathcal{H}_\mathrm{cd}^\mathrm{MF}(t) \notag \\
&=&-\frac{J}{2N}\sum_{i,j}\sigma_i^z\sigma_j^z-\Gamma(t)\sum_i\sigma_i^x \notag \\
&&-h\sum_i\sigma_i^z+\dot{\theta}(t)\sum_i\sigma_i^y. \label{Eq.adham}
\eeqs
Because this assisted Hamiltonian commutes with the square of the total spin $(\sum_i\bm{\sigma}_i)^2$, Eq.~(\ref{Eq.adham}) can be block-diagonalized. 
Therefore, in our choice of the initial state, i.e. the ground state, dynamics driven by Eq.~(\ref{Eq.adham}) is equivalent to dynamics driven by the Hamiltonian
\beq
\mathcal{H}^\mathrm{MF}(t)=-\frac{J}{S}(S^z)^2-2\Gamma(t)S^x-2hS^z+2\dot{\theta}(t)S^y,
\eeq
where $S^\alpha,\alpha=x,y,z$ is the quantum spin operator with spin-size $S=N/2$. 
Now, we consider the Schr\"odinger dynamics of this Hamiltonian
\beq
i\tder|\Psi^\mathrm{MF}(t)\rangle=\mathcal{H}^\mathrm{MF}(t)|\Psi^\mathrm{MF}(t)\rangle.
\eeq
%
%
\section{\label{Sec.Results}Results}
\subsection{\label{Sec.annealing}Quantum annealing process}
%
%
\begin{figure}
\includegraphics[width=8cm]{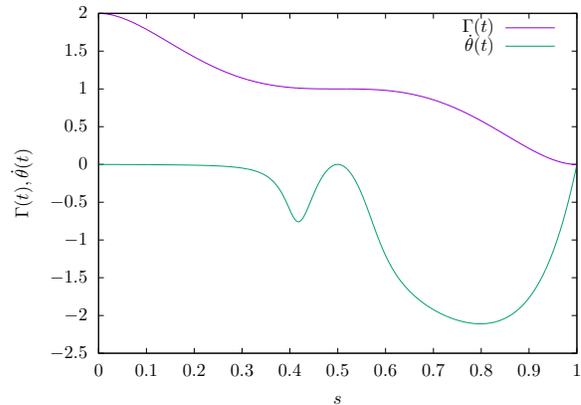}
\caption{\label{Fig.hpath} Schedule of the transverse field and the mean-field counter-diabatic field. The horizontal axis is the normalized time $s=t/t_f$ and the vertical axis is the strength of the fields $\Gamma(t)$ and $\dot{\theta}(t)$. The purple curve represents $\Gamma(t)$ and the green curve does $\dot{\theta}(t)$. }
\end{figure}
%
%

%
%
\begin{figure*}
\includegraphics[width=8cm]{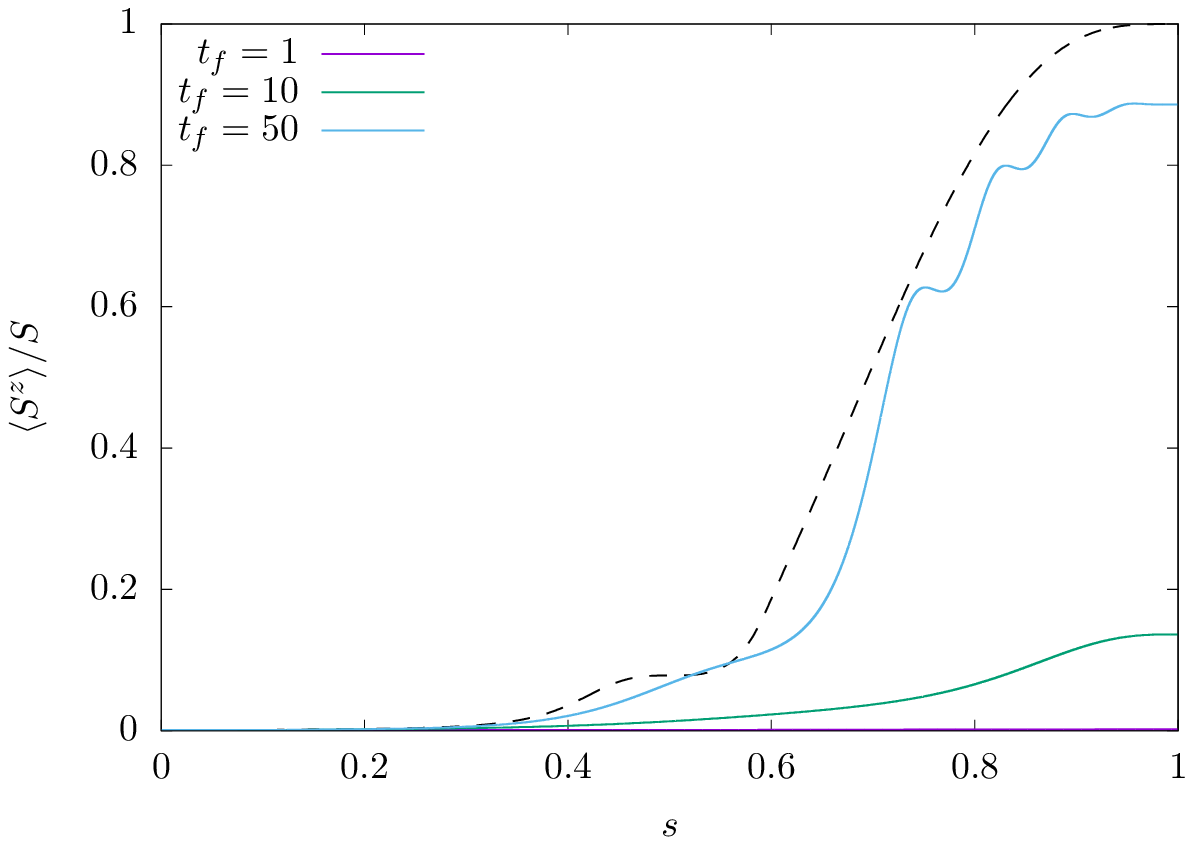}
\includegraphics[width=8cm]{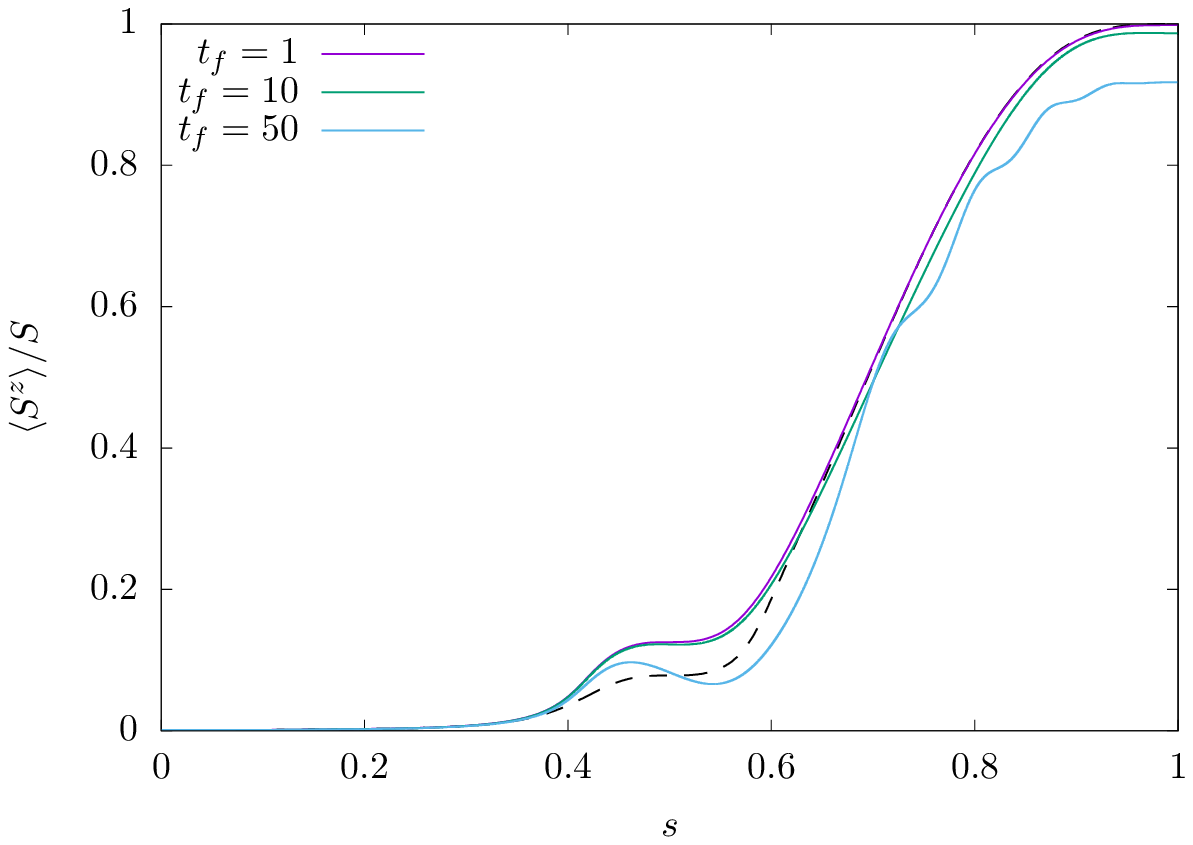}
\caption{\label{Fig.mag}Operation-time dependence of the magnetization processes (left) without and (right) with the mean-field counter-diabatic Hamiltonian. The horizontal axis is the normalized time $s=t/t_f$ and the vertical axis is normalized magnetization $\langle S^z\rangle/S$. The dotted line represents the exact adiabatic magnetization dynamics of the original Hamiltonian. The number of spins is $N=2S=1000$. }
\end{figure*}
%
%

We consider quantum annealing processes going the vicinity of the critical point, so we sweep the transverse field $\Gamma(t)$ from a positive certain value to zero and fix the longitudinal field $h$ to be infinitesimal~\cite{PhysRevE.58.5355,Suzuki2013}. 
Here, we adopt a polynomial schedule
\beqs
\Gamma(t)&=&J[48(t/t_f)^5-120(t/t_f)^4 \notag \\
&&\quad+100(t/t_f)^3-30(t/t_f)^2+2] \notag \\
&=&J[48s^5-120s^4+100s^3-30s^2+2], \label{Eq.sche}
\eeqs
where $t_f$ is the operation time and we denote $s=t/t_f$. 
Here, $\Gamma(0)=2J$, $\Gamma(t_f)=0$, and $\dot{\Gamma}(0)=\dot{\Gamma}(t_f)=0$ at the initial and final time, and this system smoothly reaches the critical point at $t=t_f/2$, $\Gamma(t_f/2)=J$ and $\dot{\Gamma}(t_f/2)=0$. 
The schedule of the transverse field~(\ref{Eq.sche}) and the consequent mean-field counter-diabatic field (\ref{Eq.cdfield}) are depicted in Fig.~\ref{Fig.hpath} for $J=1$, $h=10^{-3}$, and $t_f=1$. 
Hereafter, $J=1$ for all numerical calculations and $h=10^{-3}$ in this subsection. 

\begin{figure}
\includegraphics[width=8cm]{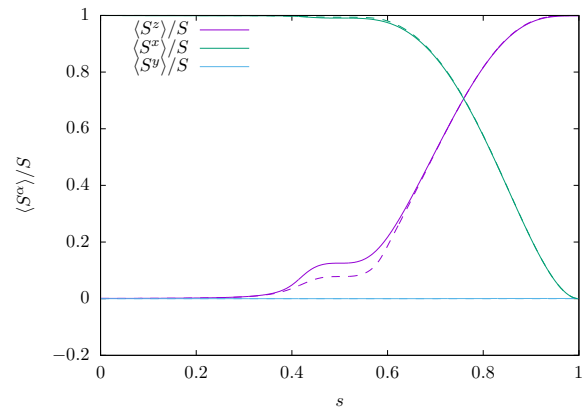}
\caption{\label{Fig.ExAdh}All the components of the magnetization dynamics. The horizontal axis is the normalized time $s=t/t_f$ and the vertical axis is the normalized magnetization $\langle S^\alpha\rangle/S, \alpha=x,y,z$. The solid curves represent the magnetization processes assisted by the mean-field counter-diabatic Hamiltonian and the dashed curves represent the exact adiabatic magnetization dynamics. Here, $N=2S=1000$ and $t_f=1$. }
\end{figure}

Figure~\ref{Fig.mag} represents the quantum annealing processes driven by (left) only the original Hamiltonian (\ref{Eq.orihamlong}) and (right) the assisted Hamiltonian (\ref{Eq.adham}). 
Here, the number of spins is $N=2S=1000$. 
With the mean-field counter-diabatic Hamiltonian, the magnetization along the $z$ axis reaches $\langle S^z\rangle/S\simeq0.998$ for the fast operation $t_f=1$, while $\langle S^z\rangle/S\simeq1.9\times10^{-3}$ without the mean-field counter-diabatic Hamiltonian. 
It is evident that the mean-field counter-diabatic Hamiltonian greatly improves adiabaticity. 
We remark that adiabaticity is not good for slow operations because errors due to the mean-field approximation are accumulated for long time. 
We also check the other components of the magnetization. 
We compare the assisted adiabatic magnetization dynamics with the exact adiabatic case in Fig.~\ref{Fig.ExAdh}. 
Here, the other components of the magnetization also have a good agreement with the exact adiabatic dynamics. 
%
%
%
%

\begin{figure*}
\includegraphics[width=8cm]{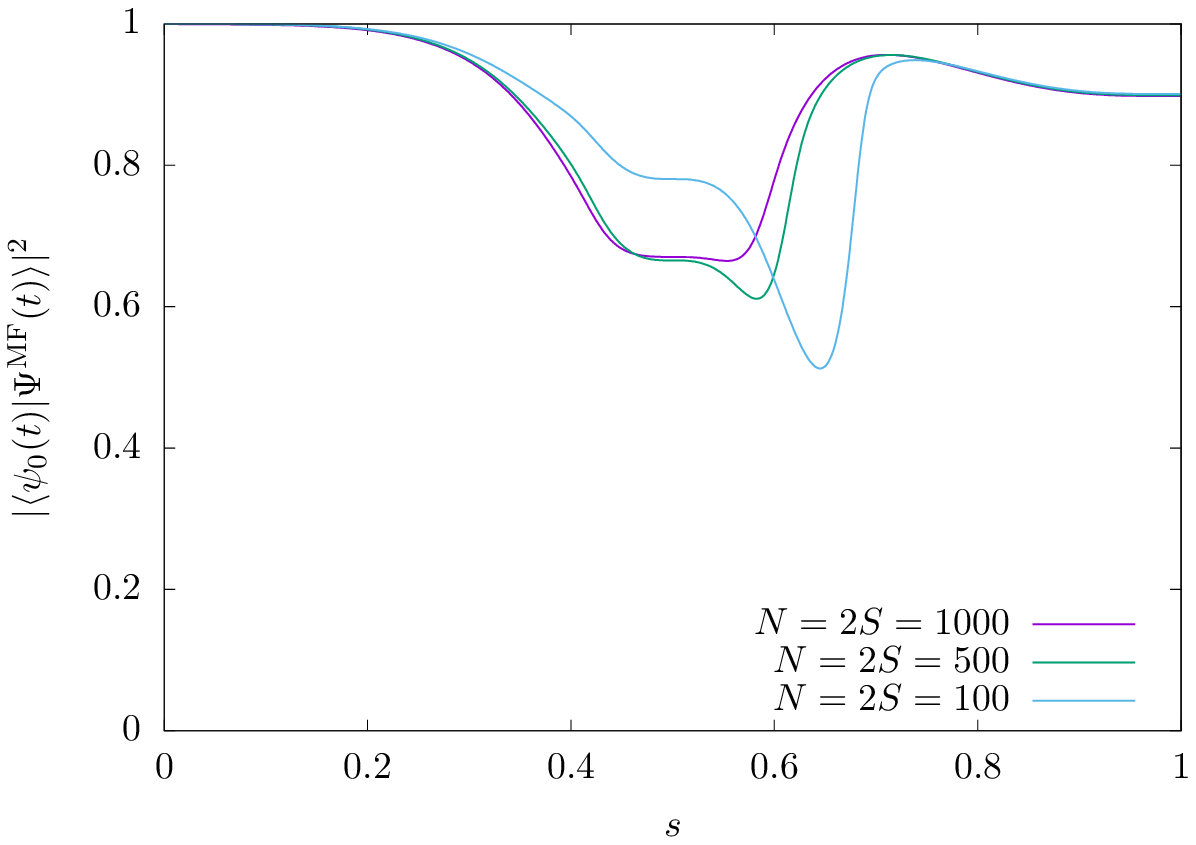}
\includegraphics[width=8cm]{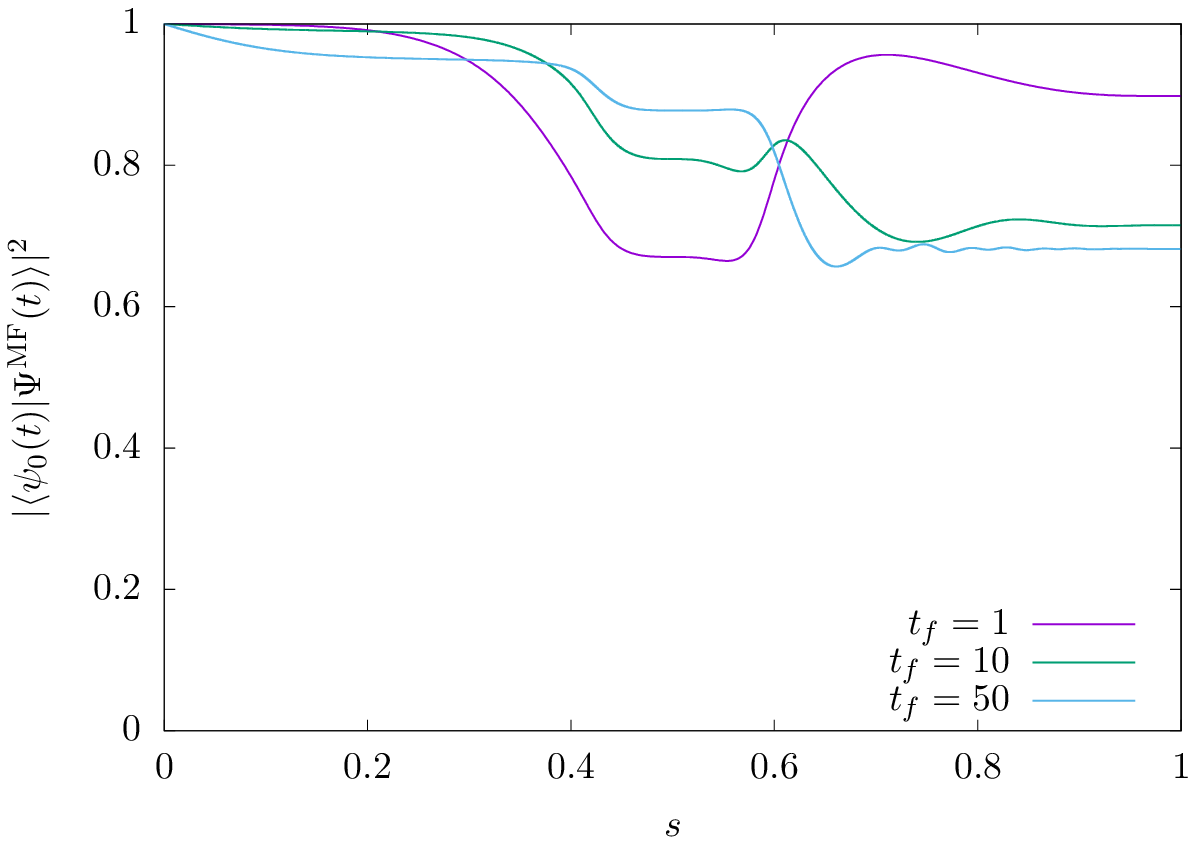}
\caption{\label{Fig.overlap}Fidelity of the adiabatic dynamics driven by the mean-field counter-diabatic field. (left) The system-size dependence of the fidelity for $t_f=1$ and (right) the operation-time dependence of the fidelity for $N=2S=1000$ are depicted. The horizontal axis is the normalized time $s=t/t_f$ and the vertical axis is the fidelity, which is the absolute square of the inner-product between the exact adiabatic state $|\psi_0(t)\rangle$ and the state driven by the mean-field counter-diabatic Hamiltonian, $|\Psi^\mathrm{MF}(t)\rangle$. }
\end{figure*}

Now, we discuss how adiabatic the system is because we cannot conclude the system is almost adiabatic just looking at the magnetization dynamics. 
Figure~\ref{Fig.overlap} represents the fidelity of the assisted adiabatic dynamics, which is the absolute square of the inner-product between the assisted state $|\Psi^\mathrm{MF}(t)\rangle$ and the exact adiabatic state $|\psi_0(t)\rangle$, i.e. $|\langle\psi_0(t)|\Psi^\mathrm{MF}(t)\rangle|^2$. 
Although there exist deviations, in particular, around the critical point, a strong deviation is removed for large $N$ (Fig.~\ref{Fig.overlap}, left). 
It should be pointed out that the fidelity in the final state is almost independent of $N$. 
It is rather surprising because, in general, it is very hard to maintain a large fidelity for a large number of spins. 
For example, if we consider the absolute square of the inner-product between a full-directed state $|\Psi_1\rangle=\otimes_{i=1}^N|\uparrow\rangle_i$ and a slightly deviated state $|\Psi_2\rangle=\otimes_{i=1}^N(\sqrt{1-\epsilon^2}|\uparrow\rangle_i+\epsilon|\downarrow\rangle_i)$ with $\epsilon\ll1$, where $|\uparrow\rangle$ and $|\downarrow\rangle$ are up and down spin states, the fidelity is suppressed as $|\langle\Psi_1|\Psi_2\rangle|^2=(1-\epsilon^2)^N\to0$ when $N\to\infty$. 
Therefore, this result tremendously supports the effectiveness of our method in the present model. 
%
%
%
%

We conclude that the mean-field counter-diabatic driving approach, which is expected to realize the quasi-adiabatic dynamics, works well and the dynamics is almost adiabatic for fast operations. 
%
%
\subsection{\label{Sec.explicit}Non-divergence \\ of the mean-field counter-diabatic field \\ in the limit to the critical point}

In the above calculations, we consider the dynamics near the critical point through the quantum annealing processes. 
Now, we show that the mean-field counter-diabatic field (\ref{Eq.cdfield}) is still well-defined in the limit $h\to+0$. 
We assume the limit $h\to+0$ in Eq.~(\ref{Eq.SCE}). 
Then, the self-consistent equation (\ref{Eq.SCE}) can be easily solved as
\beqs
m^z(t)&=&0,\quad\Gamma(t)>J, \label{Eq.mzpara} \\
m^z(t)&=&\sqrt{1-\frac{\Gamma^2(t)}{J^2}},\quad\Gamma(t)\le J. \label{Eq.mzferro}
\eeqs
Then, substituting Eqs.~(\ref{Eq.mzpara}) and (\ref{Eq.mzferro}) for Eq.~(\ref{Eq.cdfield}), we obtain the approximated expression for the mean-field counter-diabatic field
\beqs
\dot{\theta}(t)&=&0,\quad\Gamma(t)>J, \\
\dot{\theta}(t)&=&\frac{\dot{\Gamma}(t)}{2\sqrt{J^2-\Gamma^2(t)}},\quad\Gamma(t)\le J, 
\eeqs
for $h\to+0$. 
Here, the divergence appears in the mean-field counter-diabatic field $\dot{\theta}(t)$ at the critical point $\Gamma(t)=J$. 
However, this divergence can be removed by a natural choice of schedules of the transverse field $\Gamma(t)$, for example, Eq.~(\ref{Eq.sche}). 
From Eq.~(\ref{Eq.sche}), we find that the divergence of the mean-field counter-diabatic field comes from the following factor
\beq
\frac{1}{\sqrt{J^2-\Gamma^2(t)}}\propto\frac{1}{(s-1/2)^{3/2}},
\eeq
while the time derivative of the transverse field $\Gamma(t)$ produces
\beq
\dot{\Gamma}(t)\propto(s-1/2)^2. 
\eeq
Therefore, the divergence of the mean-field counter-diabatic field is removed as
\beq
\dot{\theta}(t)\propto(s-1/2)^{1/2},
\eeq
at the critical point $s=1/2$. 

\subsection{\label{Sec.local}Relation to the variational approach \\ for local counter-diabatic driving}

Sels and Polkovnikov developed the variational approach to construct the local counter-diabatic Hamiltonian~\cite{1607.05687}. 
Their method can be applied to a wide range of quantum and classical systems including many-body systems. 
It is also an advantage of their method that the eigenstates of the original Hamiltonian are not necessary to construct the counter-diabatic Hamiltonian. 
In this subsection, we explain the relationship between their and our methods. 

Regarding our Hamiltonian (\ref{Eq.orihamlong}), we require the following form of the gauge potential
\beq
\mathcal{A}^\ast(t)=\alpha(t)\sum_i\sigma_i^y,
\label{Eq.gauge}
\eeq
where $\alpha(t)$ is nothing but the counter-diabatic field $\dot{\theta}(t)$ in the present paper. 
From Eqs.~(\ref{Eq.orihamlong}) and (\ref{Eq.gauge}), we obtain the following function
\beqs
G(t)&\equiv&\partial_t\mathcal{H}_0(t)+i[\mathcal{A}^\ast(t),\mathcal{H}_0(t)] \notag \\
&=&-(\dot{\Gamma}(t)-2h\alpha(t))\sum_i\sigma_i^x \notag \\
&&+\frac{J\alpha(t)}{N}\sum_{i,j}(\sigma_i^z\sigma_j^x+\sigma_i^x\sigma_j^z) \notag \\
&&-2\Gamma(t)\alpha(t)\sum_i\sigma_i^z, 
\eeqs
or equivalently
\beqs
G(t)&=&-2(\dot{\Gamma}(t)-2h\alpha(t))S^x \notag \\ 
&&+\frac{2J\alpha(t)}{S}(S^zS^x+S^xS^z) \notag \\
&&-4\Gamma(t)\alpha(t)S^z, 
\eeqs
as discussed in Sec.~\ref{Sec.numsim}. 
We calculate the Hilbert-Schmidt norm of this function and obtain
\beqs
\mathrm{Tr}G^2(t)&=&\frac{4}{3}(\dot{\Gamma}(t)-2h\alpha(t))^2S(S+1)(2S+1) \notag \\
&&+\frac{1}{15}\left(\frac{2J\alpha(t)}{S}\right)^2S(S+1)(2S+1) \notag \\ 
&&\hspace{3cm}\times(2S-1)(2S+3) \notag \\
&&+\frac{1}{3}(4\Gamma(t)\alpha(t))^2S(S+1)(2S+1). 
\eeqs
By minimizing this norm with respect to $\alpha(t)$, we obtain the local counter-diabatic field
\beq
\alpha(t)=\frac{1}{2}\frac{h\dot{\Gamma}(t)}{h^2+\Gamma^2(t)+(J/S)^2(2S-1)(2S+3)/20}, 
\eeq
where the correction factor $(J/S)^2(2S-1)(2S+3)/20$ vanishes for $J=0$ or $S=1/2$, for which the long-range Ising model is equivalent to the independent $N$ two-level spins. 

It is obvious that this local counter-diabatic field does not equal to the mean-field counterdiabatic field~(\ref{Eq.cdfield}). 
If we adopt the mean-field Hamiltonian (\ref{Eq.LongMFham}) instead of (\ref{Eq.orihamlong}), we can obtain the mean-field counter-diabatic field (\ref{Eq.cdfield}) as the local counter-diabatic field although there is no longer an advantage of the variational approach, which is the unnecessity of the eigenstates. 
This result suggest that our mean-field counter-diabatic driving approach takes higher order contributions into account and is consistent with the variational approach for local counter-diabatic driving. 

%
%
\section{\label{Sec.Summary}Summary}
In this article, we constructed the counter-diabatic Hamiltonian for the infinite-range Ising model (\ref{Eq.orihamlong}) by making use of the mean-field approximation. 
This mean-field counter-diabatic Hamiltonian (\ref{Eq.cdhamlong}) is quite useful because it is constructed by only the local operators. 
However, owing to the mean-field approximation, we had to check if this mean-field counter-diabatic Hamiltonian can assist the adiabatic dynamics of the original Hamiltonian. 

First, we tested the mean-field counter-diabatic driving in the quantum annealing processes. 
If we adopt the polynomial schedule (\ref{Eq.sche}), the mean-field counter-diabatic field (\ref{Eq.cdfield}) becomes a two-pulse like shape as seen in Fig.~\ref{Fig.hpath}. 
It was found that the quasi-adiabatic dynamics realizes under the mean-field counter-diabatic field (\ref{Eq.cdfield}) as seen in Figs.~\ref{Fig.mag}, \ref{Fig.ExAdh}, and \ref{Fig.overlap}. 
In particular, it was the surprising result that the fidelity of the final state is almost independent of the system size $N$. 
We also confirmed the non-divergence of the mean-field counter-diabatic field at the critical point although we require an infinitesimal longitudinal field $h=+0$ in order to break the symmetry. 

We also investigated the relation to the variational approach for local counter-diabatic driving~\cite{1607.05687}. 
We found that our mean-field counter-diabatic driving approach takes higher order contributions into account. 
This contributions cannot be taken into account if we naively adopt the variational approach. 
If we apply the mean-field approximation to the variational approach, we can obtain the same local counter-diabatic field in the present paper. 
However, this treatment spoils the unnecessity of the eigenstates, which is an advantage of the variational approach. 

Finally, we discuss application of the mean-field counter-diabatic driving approach. 
One of the application which we can immediately apply to is fast magnetization switching. 
When we consider magnetization reversal of uniaxial single-spin magnets under finite-rate magnetic field sweeping, systems are inevitable to be excited associated with the formation of the hysteresis caused by the first order phase transitions~\cite{PhysRevLett.116.037203,1704.06466}. 
The mean-field counter-diabatic driving approach can avoid such excitations. 
On the other hand, there exist gaps to realize the quantum annealing processes in experiments, which we demonstrated in this paper. 
We use the information of the ground state to calculate the mean-field. 
However, the purpose of quantum annealing is nothing but to obtain the information of the ground state. 
Therefore, this method cannot apply to quantum annealing in the present formalism. 
Further development in order to apply to quantum annealing should be investigated in the future. 
%
%
\begin{acknowledgments}
The author is grateful to Dr. Takashi Mori and Prof. Seiji Miyashita for fruitful comments and also thanks the anonymous referee for useful comments. 
This work is supported by the Ministry of Education, Culture, Sports, Science and Technology (MEXT) of Japan through the Elements Strategy Initiative Center for Magnetic Materials. 
The author is supported by the Japan Society for the Promotion of Science (JSPS) through the Program for Leading Graduate Schools: Material Education program for the future leaders in Research, Industry, and Technology (MERIT) of the University of Tokyo. 
\end{acknowledgments}
%
%
\bibliography{mfsbib}

\end{document}